\documentclass[prd,amsmath,
twocolumn,floatfix,amssymb, preprintnumbers, nofootinbib, superscriptaddress]{revtex4}

\usepackage{amsmath,amssymb,amsfonts}
\usepackage{graphicx}
\usepackage{hyperref}
\usepackage{slashed}
\usepackage{booktabs,tabulary}
\usepackage{mciteplus}
\usepackage{color}
\usepackage{physics}
\usepackage[percent]{overpic}
\usepackage{soul}

\usepackage{bibentry}
\newcommand{\ignore}[1]{}
\newcommand{\nobibentry}[1]{{\let\nocite\ignore\bibentry{#1}}}

\makeatletter
\def\bibinfo@X@title#1,{\ignorespaces}
\makeatother 

\begin{document}

\title{Simultaneous description of the $e^+e^- \to J/ \psi \, \pi \pi\, (K \bar{K})$ processes}

\author{Igor Danilkin}
\author{Daniel A. S. Molnar}
\author{Marc Vanderhaeghen}
\affiliation{Institut f\"ur Kernphysik \& PRISMA$^+$  Cluster of Excellence, Johannes Gutenberg Universit\"at,  D-55099 Mainz, Germany}

\date{\today}

\begin{abstract}
In this work, we provide a simultaneous and accurate description of the $\pi^+\pi^-$ and $\pi^{\pm} J/\psi$ invariant mass distributions of the recent BESIII data on  $e^+ e^- \to J/\psi \; \pi^+ \pi^-$ together with the $e^+ e^- \to J/\psi \; K^+K^-$ cross sections at $e^+e^-$ center-of-mass energies $q=4.23$ GeV and $q=4.26$ GeV. The rescattering effects between pions in the S and D-waves are taken into account through the Muskhelishvili-Omn\`es formalism. Since the physical region of the $\pi\pi$ invariant mass extends above 1 GeV, the important $K\bar{K}$ intermediate state in the S-wave is implemented through coupled-channel unitarity. For the left-hand cuts, we account for the well established charged exotic state $Z_c(3900)$ in $t$- and $u$-channels, while the other contributions are absorbed in the subtraction constants. For the $e^+ e^- \to J/\psi\, K \bar{K}$ we provide the prediction of the two-kaon invariant mass distribution. The constructed amplitudes serve as an essential framework to interpret the present and forthcoming measurements by the BESIII and Belle II Collaborations.
\end{abstract}

\maketitle

\section{Introduction}\label{intro}
The charged exotic charmonium-like state $Z_c(3900)$ was discovered simultaneously by the BESIII and Belle Collaborations in 2013 both in direct production \cite{Ablikim:2013mio} and using initial-state radiation \cite{Liu:2013dau}, in the process $e^+ e^- \to J/\psi \pi^+ \pi^-$ and soon confirmed using the CLEO-c data~\cite{Xiao:2013iha}. In 2015, the neutral partner was observed by the BESIII Collaboration in the same reaction with neutral pions $e^+ e^- \to J/\psi \pi^0 \pi^0$ \cite{Ablikim:2015tbp}. Recently, the D0 Collaboration, using  proton-antiproton collisions, has found a signal of $Z_c(3900)$ in non-prompt semi-inclusive weak decays of b-flavored hadrons \cite{D0:2019zpb,Abazov:2018cyu}. Furthermore, in recent years, BESIII has observed $Z_c(3900)$ in the $e^+ e^- \to (D\bar{D}^*)^{\mp} \pi^\pm$ process using a single-tag analysis \cite{Ablikim:2013xfr}, a double-tag analysis \cite{Ablikim:2015swa}, and also by analyzing the neutral channel $e^+ e^- \to (D\bar{D}^*)^{0} \pi^0$ \cite{Ablikim:2015gda}. The most precise data so far has been reported in Ref.\cite{Collaboration:2017njt}, where an updated BESIII analysis of $e^+ e^- \to J/\psi \pi^+ \pi^-$ allow us to determine the spin-parity $J^P = 1^+$ assignment of the $Z_c(3900)$. 

From the theory side, the nature of $Z_c(3900)$ is still a puzzle \cite{Guo:2019twa,Liu:2019zoy, Guo:2017jvc,Esposito:2016noz}. Most likely it corresponds to a pole in the unphysical Riemann sheet, which could be a hadro-charmonium \cite{Dubynskiy:2008mq,Danilkin:2011sh}, molecular state \cite{Guo:2017jvc,Guo:2014iya} or a virtual state \cite{Albaladejo:2015lob,Ortega:2018cnm}. The peak at the $Z_c(3900)$ position has also been interpreted through a kinematic effect \cite{Wang:2013cya,Liu:2013vfa,Szczepaniak:2015eza}. The most popular scenarios correspond to the triangle singularity associated with $D^*_0(2300)\bar{D}^*D$ \cite{Liu:2013vfa} or $D_1(2420)\bar{D}^*D$ \cite{Szczepaniak:2015eza} loops. In both cases, the left-hand cut branch point stays relatively far away from the physical region, either due to the large width of $D^*_0(2300)$, or due to the off-shellness of $D_1(2420)$ for $q=4.23$ GeV and $q = 4.26$ GeV, and only the $\bar{D}^*D$ threshold cusp gets enhanced. However, as it was pointed out in Ref.\cite{Gong:2016jzb}, the recent BESIII data  \cite{Collaboration:2017njt} indicate that the $Z_c(3900)$ peak is more enhanced for $q=4.23$ GeV compared to $q=4.26$ GeV, in contrast to what one  expects from the threshold cusp enhancement mechanism due to the triangle singularity. Additionally, the contribution from the rescattering process has to be accounted for, which typically smooths out kinematic singularities. To shed further light on this puzzle, it will be very helpful to observe the $Z_c(3900)$ in other decay modes \cite{Brambilla:2019esw}. Besides, it is important to clarify if there exists a possible strange partner of $Z_c(3900)$, the so-called $Z_{cs}$,  which can show up in the $K J/\psi$ distribution of the $e^+ e^- \to J/\psi K^+ K^-$ process. So far, Belle \cite{Shen:2014gdm} and BESIII \cite{Ablikim:2018epj} have not seen a clear structure in the $K J/\psi$ mass distribution, and future high statistics measurements are necessary.

The purpose of the present work is to demonstrate a dispersive amplitude analysis, which can be applied in the experimental works to describe the whole Dalitz plot with minimum assumptions about the nature of the charged $Z_c$ state. Our work is a continuation of the previous work \cite{Molnar:2019uos}, where for the first time, a dispersive amplitude analysis was applied to describe $e^+e^-\to \psi(2S)\pi^+\pi^-$ Dalitz plot projections \cite{Ablikim:2017oaf,Ablikim:2017aji}. In our current analysis, the recent BESIII \cite{Collaboration:2017njt} data on $e^+ e^- \to J/\psi \pi^+ \pi^-$ play the central role. We present a simultaneous description of the $\pi^+\pi^-$ and $\pi^{\pm} J/\psi$ invariant mass distributions by providing rigorous dispersive treatment of the $\pi\pi$ final state interactions. We account for $Z_c(3900)$ as an explicit degree of freedom in the $t$- and $u$-channels and unitarize the $\pi\pi$ final state interaction on the base of the Muskhelishvili-Omn\`es formalism. Other possible left-hand cut contributions are absorbed in the subtraction constants which we determine from a combined fit to the $e^+ e^- \to J/\psi \pi^+ \pi^-$ Dalitz plot data and the total cross-section data for $e^+ e^- \to J/\psi K^+K^-$. Due to the relatively large physical region of the $\pi\pi$ invariant mass, we also extend our previous analysis of \cite{Molnar:2019uos} to the coupled-channel in the $\pi \pi$ S-wave and include the D-wave.  Allowing for a minimum number of parameters, which enter in the form of subtraction constants, and assuming the absence of $Z_{cs}$ at $q=4.23$ GeV and $q = 4.26$ GeV\footnote{Due to the strange quark mass, it is reasonable to assume that the strange partner of $Z_c(3900)$ would have a heavier mass (in particularly, Ref. \cite{Lee:2008uy} predicts a mass of $3.97\pm 0.08$ GeV) and therefore $Z_{cs}$ cannot be seen as peak in the $K\bar{K}$ invariant mass distribution for $q=4.23 - 4.26$ GeV.}, we provide a prediction for the invariant mass distribution for the process $e^+ e^- \to J/\psi \; K^+ K^-$.

In our analysis we do not aim at a description of the full $e^+ e^- \to J/\psi \pi^+ \pi^-$ cross-section and instead implement the $q^2$ dependence model independently, by applying our formalism for each $q$-value independently. The study of the two possible resonance structures seen in the $e^+ e^- \to J/\psi \pi^+ \pi^-$ total cross-section \cite{Ablikim:2016qzw} is beyond the scope of this paper. Rather, we want to use the available Dalitz plot projection data to make a simultaneous description of both $\pi^+\pi^-$ and $\pi^{\pm} J/\psi$ invariant mass distributions and obtain a prediction of the $K^+K^-$ and $K^{\pm} J/\psi$ invariant mass distributions. This is different from the analysis performed in Ref.\cite{Chen:2019mgp}, which focused only on the $\pi^+\pi^-$ invariant mass distribution to get insights into the structure of the $Y(4260)$ state from the light-quark perspective. Though the analysis of the $\pi\pi$ final state interaction is similar in spirit to ours, there are several technical differences, which we will point out below.

\section{Kinematics}
The double differential cross section for the $e^-(p_1)\, e^+(p_2) \to \gamma^* (p_{\gamma^*})\to J/\psi(p_{\psi})\, \pi^+(p_{\pi^+})\, \pi^- (p_{\pi^-})$ process can be written as
\begin{equation}
\frac{d ^2 \sigma}{d s\, d t} =  \frac{e^2}{2^5 (2 \pi)^3 \, q^6} 
\frac{1}{3} \;
\left[
\displaystyle \sum_{\lambda_1 \lambda_2}   
|\mathcal{H}_{\lambda_1 \lambda_2}|^2
\right],
\label{cross-section}
\end{equation}
where we have neglected the electron mass compared to the $e^+e^-$ center of mass (CM) energy $q=\sqrt{p_{\gamma^*}^2}$. In Eq.(\ref{cross-section}) the helicity amplitudes $\mathcal{H}_{\lambda_1 \lambda_2}$ are defined in the usual way,
\begin{align}
&\bra{\pi \pi \psi(\lambda_2)} \mathcal{T} \ket{\gamma^{*}(\lambda_1)}\\ 
&\quad = (2 \pi)^4
\,\delta(p_{\gamma^*} - p_\psi - p_{\pi^+} - p_{\pi^-}) \; \mathcal{H}_{\lambda_1\lambda_2},\nonumber
\end{align}
with
\begin{align}
\mathcal{H}_{\lambda_1\lambda_2}\equiv &\,\mathcal{H}^{\mu\nu}\epsilon_{\mu}(p_{\gamma^*},\lambda_1)\, \epsilon_{\nu}^{*}(p_{\psi},\lambda_2)\, ,
\end{align} 
where $\lambda_1(\lambda_2)$ denote the $\gamma^* (J/\psi)$ helicities, respectively. For the process $ \gamma^* \to J/\psi\, \pi^+\pi^-$ the following Mandelstam variables are chosen,
\begin{align}
&s = (p_{\pi^+} + p_{\pi^-})^2 \equiv M^2_{\pi^+\pi^-}\,,\nonumber \\ 
&t =(p_{\psi} + p_{\pi^-})^2\equiv 
M^2_{\pi^-\psi}\,,\\
& u = (p_{\psi} + p_{\pi^+})^2\equiv M^2_{\pi^+\psi} ,
\nonumber
\end{align}
which satisfy
\begin{align}
&s +  t + u = q^2 + m_{\psi}^2 + 2 m_{\pi}^2\,.
\end{align}
In the following we use the kinematics in the CM frame of the di-pion system, and define $z \equiv \cos \theta_s $ as the cosine of the angle between the $p_{\pi^+}$ and the $p_\psi$ momenta. Thus, in this frame the following relations hold
\begin{align}
t(s, z) = \frac{1}{2}(q^2+m_{\psi}^2 + 2 m_{\pi}^2 - s) + \frac{\kappa(s)}{2}\, z,
\nonumber \\
u(s, z) = \frac{1}{2}(q^2+m_{\psi}^2 + 2 m_{\pi}^2 - s) - \frac{\kappa(s)}{2}\, z,
\end{align}
where 
\begin{align}
\kappa(s) = \frac{1}{s} \sqrt{\lambda(s,q^2,m_{\psi}^2)\, \lambda(s,m_{\pi}^2,m_{\pi}^2)} ,
\end{align}
and $\lambda$
being the K\"allen function. Consequently, $z$ can be written in terms of $t$ and $u$ as
\begin{align}\label{Eq:z}
z = \dfrac{t- u}{\kappa(s)} .
\end{align}

\section{Dispersive Formalism}\label{dispersionR}
In this section, we briefly describe the dispersive formalism that we adopt to account for the rescattering between two pions (kaons), which generates the most important singularities at low energies in the $s$-channel. The partial wave (p.w.) expansion reads
\begin{align}
\mathcal{H}_{I,\lambda_1 \lambda_2}(s,t) = \sum\limits_{J=0}^{\infty} (2J + 1)\, d^{(J)}_{\Lambda,0}(\theta_s)\, h^{(J)}_{I,\lambda_1 \lambda_2}(s)\,,
    \label{Hdef}
\end{align}
where $I$ is the isospin, $\Lambda = \lambda_1-\lambda_2$ and $d^{(J)}_{\Lambda,0}$ is the Wigner rotation function. For better readability, below we will consistently suppress the isospin indices, and retrieve them at the beginning of Sec.\ref{results}. On account of causality, the p.w. amplitudes should have contributions from the left- and right-hand cuts,
\begin{align}\label{Eq:Split}
h_{\lambda_1 \lambda_2}^{(J)}(s) = h^{(J),L}_{\lambda_1 \lambda_2}(s)+ h^{(J),R}_{\lambda_1 \lambda_2}(s)\,,
\end{align}
where the branch cut due to the two-pion interaction starts at $s=4\,m_\pi^2$. We note that the amplitudes $ h_{\lambda_1 \lambda_2}^{(J)}(s)$ are subject to kinematical constraints, which in principle have to be removed before application of dispersion relations.  The hadron tensor $\mathcal{H}^{\mu\nu}$ of $\gamma^* \to J/\psi \,  \pi\pi$ can be decomposed into a suitable set of Lorentz structures given in Ref.\cite{Molnar:2019uos} (see also \cite{Tarrach:1975tu,Drechsel:1997xv,Colangelo:2015ama,Hoferichter:2019nlq,Danilkin:2019opj}),
\begin{align}
\mathcal{H}^{\mu \nu} = \displaystyle\sum_{i= 1}^{5} F_i L_i^{\mu \nu},
\end{align}
with $F_i$ the corresponding invariant amplitudes.  
One can then show that for the S-wave the p.w. helicity amplitudes are correlated at the kinematic points  $s=(q \pm m_\psi)^2$,
\begin{align}\label{KinConstr}
h_{++}^{(0)}(s) \pm h_{00}^{(0)}(s)\sim {\cal O}(s-(q \pm m_\psi)^2)\,,
\end{align}
while for the D-wave the kinematic correlations between different p.w. helicity amplitudes are more complicated and can be found in Ref.\cite{Danilkin:2019opj}. As it will be shown in the next section, for the considered kinematics most of these constraints have a negligible impact on the results, since the sum in Eq.(\ref{cross-section}) in the physical region can be written in terms of $\mathcal{H}_{++}$ only, i.e.
\begin{align}
\displaystyle\sum_{\lambda_1\lambda_2}  |\mathcal{H}_{\lambda_1\lambda_2}|^2  \approx  3\, |\mathcal{H}_{++}|^2\,.
\label{Hsquared}
\end{align}
Under this approximation it is enough to take into account only the so-called centrifugal barrier factor for $J=2$
\begin{align}\label{KinConstr2}
&h_{++}^{(2)}(s)\sim {\cal O}\left(\gamma(s)\right)\,,\\
&\gamma(s)\equiv (s-4m_\pi^2)(s-(q-m_\psi)^2)\,,\nonumber
\end{align}
which comes from the properties of the Legendre polynomials entering p.w. expansion in Eq.(\ref{Hdef}). We note, however, while Eq.(\ref{KinConstr2}) is exact for $s=4m_\pi^2$, a zero at $s=(q-m_\psi)^2$ is only approximate and typically a few MeV away. This is related to the approximation made in Eq.(\ref{Hsquared}), which we will discuss further on.

The discontinuity across the branch cut in the s-channel is given by
\begin{align}\label{Eq:Disc}
\text{Disc}\, h_{++}^{(J)}(s)&=\frac{1}{2\,i}(h_{++}^{(J)}(s+i\,\epsilon )-h_{++}^{(J)}(s-i\,\epsilon))\nonumber\\
& = t^{(J)*}(s)\,\rho(s)\,h_{++}^{(J)}(s)\,,
\end{align}
which can be straightforwardly extended to the case of two cuts (coupled-channel case) in the S-wave
\begin{align}\label{Eq:DiscCC}
\begin{bmatrix}
\text{Disc} \; h_{++}^{(0)}(s)
\\[0.25cm]
\text{Disc} \; k_{++}^{(0)}(s)
\end{bmatrix}
= t^{(0)*}(s)\,\rho(s)
\begin{bmatrix}
 h_{++}^{(0)}(s)
\\[0.25cm]
 k_{++}^{(0)}(s)
\end{bmatrix}\,.
\end{align}
The two-body phase space $\rho(s)$ is given by
\begin{align}
\rho(s) =
\frac{1}{16 \,\pi}
\begin{bmatrix}
\sigma_{\pi\pi}\,  \theta(s-4 m_\pi^2) & 0
\\[0.25cm]
0 & \sigma_{KK}\,  \theta(s-4 m_K^2)
\end{bmatrix}\,,
\end{align}
where $\sigma_{\alpha \alpha}(s)=\lambda^{1/2}(s,m_\alpha^2,m_\alpha^2)/s$, with $\alpha = \pi$ or $K$.
The $\{\pi\pi, K\bar{K} \}$ coupled-channel scattering amplitude $t(s)$ is normalized as $\text{Disc}\,(t^{(0)}(s))^{-1}=-\rho(s)$. In Eq.(\ref{Eq:DiscCC}), $k_{\lambda_1 \lambda_2}^{(0)}(s)$ is the S-wave amplitude of the total helicity amplitude $\mathcal{K}_{++} (s,t)$ for $\gamma^*(q) \to J/\psi \, K\bar{K}$. We note, that in the p.w. expansion of the $\gamma^* (q)\to J/\psi \, K\bar{K}$ process we include an extra factor $1/\sqrt{2}$ in contrast to $\gamma^*(q) \to J/\psi \, \pi\pi$ in order to match our normalization for the hadronic p.w. amplitudes, which ensure the same unitarity relations for the identical and non-identical particles. 
For the S-wave the standard Muskhelishvili-Omn\`es representation
for the left-hand cut subtracted p.w. amplitude is given
by (modulo subtractions)
\begin{align}\label{Eq:RescSwave}
\begin{bmatrix}
h^{(0),R}_{++}
\\[0.2cm]
k^{(0),R }_{++}
\end{bmatrix} 
&=  -\Omega^{(0)}
 \int\limits_{4m_\pi^2}^{\infty} 
 \dfrac{d s'}{\pi} \dfrac{\text{Disc}\, (\Omega^{(0)}(s'))^{-1}}{s' - s} 
 \begin{bmatrix}
h^{(0),L}_{++}(s')
\\[0.2cm]
k^{(0),L}_{++}(s')
\end{bmatrix},
\end{align}
where the coupled-channel Omn\`es function (with $1=\pi\pi$ and $2=K\bar{K}$)
\begin{align}
\Omega^{(0)}(s)=\begin{bmatrix}
\Omega^{(0)}_{11}(s) & \Omega^{(0)}_{12} (s)
\\[0.2cm]
\Omega^{(0)}_{21}(s) & \Omega^{(0)}_{22}(s) 
\end{bmatrix},
\end{align}
satisfies the following unitarity relation
\begin{align}\label{Eq:Omnes}
\text{Disc} \,\Omega^{(J)}(s) &= t^{(J)*}(s)\,\rho(s) \,\Omega^{(J)}(s)\,.
\end{align}
Since the tail of the $f_2(1270)$ resonance could overlap with the physical region, we include D-wave single-channel $\pi\pi$-rescattering. As discussed previously, we factor out the known threshold factor and write a dispersion relation for $h^{(2),R}_{++}(s)\,(\Omega^{(2)}(s))^{-1}/\gamma(s)$ which leads to
\begin{align}\label{Eq:RescDwave}
h_{++}^{(2),R}(s)&=\gamma(s)\,\Omega^{(2)}(s)\\
&\times \bigg\{- \int\limits_{4m_\pi^2}^{\infty}  \frac{ds'}{\pi}\frac{\text{Disc}\, (\Omega^{(2)}(s'))^{-1}}{(s'-s)}\frac{h^{(2),L}_{++}(s')}{\gamma(s')}\bigg\}\,,\nonumber
\end{align}
where under the dispersive integral we slightly adjusted a zero of $\gamma(s')$ at $s'=(q-m_\psi)^2$  to match exactly a zero of $h^{(2),L}_{++}(s')$, which is few MeV away.
One can notice, that the overall threshold factor $\gamma(s)$ is also needed to compensate the singularities of $z=\cos\theta_s$ (see Eq.(\ref{Eq:z})) of the full amplitude $\mathcal{H}^{R}_{++}(s,t) $ at the borders of the Dalitz plot (i.e. at $s=4m_\pi^2$ and $s=(q-m_\psi)^2$). This is different from Ref.\cite{Chen:2019mgp} where in the dispersive representation no threshold factors were taken into account in the D-wave.

In our formalism, we are accounting for the $\pi\pi$ rescattering effects only in S- and D-waves, and beyond that (for $J>2$) the p.w. amplitudes in Eq.(\ref{Eq:Split}) are approximated by the first term, $h^{(J),L}_{\lambda_1\lambda_2}(s)$. In other words, we keep the cross channel p.w. expansion to all orders. That is crucial to get the description of the full Dalitz plot, where there are peaks structures in both $\pi\pi$ and $\pi J/\psi$ systems. The final result for the total helicity amplitude can be written as\footnote{We note the difference between Eq.(\ref{Hdef2}) and the reconstruction theorem written in Ref.\cite{Molnar:2019uos}. The latter is correct only for the scalar particles and needs to be modified for the particles with spin Ref.\cite{Albaladejo:2019huw}. Since we only considered rescattering effects in the $s$-channel and $h_{++}^{(0),t}(t)+h_{++}^{(0),u}(u)$ almost coincides with $\mathcal{H}_{++}^{L}(s,t)$ in Eq.(\ref{Hdef2}), this has no effect on the results in Ref.\cite{Molnar:2019uos}. }
\begin{align}
\mathcal{H}_{++}(s,t) = \mathcal{H}_{++}^{L} (s,t) + \sum\limits_{J=0}^{2} (2J + 1)\, P_{J}(z)\, h^{(J),R}_{++}(s),
\label{Hdef2}
\end{align}
where the sum goes only over even $J$ values due to Bose symmetry of two pions and C-parity conservation.

\subsection{Left-hand cuts}
The cuts associated with the crossed channel exchange terms, i.e. $h^{(J),L}_{\lambda_1 \lambda_2}(s)$,  are approximated by the charged $Z_c$ exchanges, motivated by the experimental data \cite{Collaboration:2017njt}, where the $Z_c(3900)$ axial-vector state and its kinematic reflection show up as clear peaks in the $\pi J/\psi$ projection for both $e^+e^-$-CM energies $q=4.23$ GeV and $q = 4.26$ GeV. According to the mechanism  $\gamma^*(q^2) \to \pi^{\mp} + (Z_{c}^{\pm} \to  J/ \psi + \pi^{\pm}) $, the helicity amplitude can be expressed in a general form as follows 
\begin{align}
\mathcal{H}^{Z_c}_{\lambda_1\lambda_2} &= (V_{Z_{c} \psi \pi})^{\beta \nu} \, 
S_{\nu \mu}(Q_{z}) \,
(V_{\gamma^* \pi Z_{c} })^{\mu \alpha}  \,\nonumber\\
&
\qquad \times\epsilon_{\alpha}(p_{\gamma^*},\lambda_1)\, \epsilon_{\beta}^{*}(p_{\psi},\lambda_2) ,
\label{Mc}
\end{align}
where $S_{\nu \mu}(Q_z)$ is the axial meson propagator. We use the following vertices \cite{Roca:2003uk}
\footnote{In general there are two vertex structures for the axial-vector-pseudoscalar transition. The different choices used in the literature were e.g. reviewed 
in Ref.\cite{Lichard:2006kw}. As we only need the on-shell vertices for our purpose, we can conveniently choose the second  vertex structure of the form:
\begin{align}
(V^{(2)}_{Z_{c} \psi \pi})^{\beta \nu} &= C_{2} \;  p_{\pi}^{\nu}
\; \left( Q_{z}^{\beta} - \frac{p_\psi \cdot Q_z}{p_\psi^2} p_\psi^\beta \right),\nonumber
\end{align}
and an analogous expression for the  second 
$V_{\gamma^* \pi Z_{c} }^{(2)}$ vertex. 
We checked that by including the second vertex structures with the same order of magnitude of the couplings, only leads to a very small difference for the total unpolarized  result. One reason for the small relative contribution with the above choice of vertex structure 2 is the suppression due to the pion four-momentum.  
Therefore for the purpose of the unpolarized observable, the use of one effective coupling (vertices in Eq.(\ref{Vz+psipi})) can be applied and its value adjusted accordingly. 
}, 
\begin{align}
(V_{ Z_{c} \psi \pi})^{\beta \nu} &= C_{Z_{c} \psi \pi} \;  \left( g^{\beta \nu} \left( p_{\psi} \cdot Q_{z} \right) - p_{\psi}^{\nu} Q_{z}^{\beta} \right) ,
\label{Vz+psipi}
\\
(V_{\gamma^* \pi Z_{c}} )^{\mu \alpha} &= \mathcal{F}_{\gamma^* \pi Z_{c}}(q^2) \; \left( g^{\alpha \mu} \left( p_{\gamma^*} \cdot Q_{z}\right)  - p_{\gamma^*}^{\mu} Q_{z}^{\alpha} \right) ,
\nonumber 
\end{align}
where $Q_{z} = (p_{\gamma^*} - p_{\pi})$, $C_{Z_{c} \psi \pi}$ the coupling among $Z_c$, $J/\psi$ and $\pi$ and $\mathcal{F}_{\gamma^* \pi Z_{c}}(q^2)$ is the corresponding transition form factor. For the present analysis, the latter should in principle encode for two resonances, as observed in the data \cite{Ablikim:2016qzw}. In our formalism we will perform two independent analyses at $q=4.23$ GeV and $q=4.26$ GeV, without any assumption for $\mathcal{F}_{\gamma^* \pi Z_{c}}(q^2)$ to avoid possible model dependence.

By inspecting Eq.\eqref{Mc} for our particular kinematics, we observe that the helicity amplitudes, $\mathcal{H}_{++}^{Z_c}$ and $\mathcal{H}_{00}^{Z_c}$ give the main contribution compared to other helicity amplitudes. Furthermore, $\mathcal{H}_{++}^{Z_c}$ and $\mathcal{H}_{00}^{Z_c}$ turn out to be numerically very close to each other $|\mathcal{H}_{++}^{Z_c}| \approx |\mathcal{H}_{00}^{Z_c}|$. Therefore, one can write the sum in Eq.(\ref{cross-section}) in terms of only $\mathcal{H}_{++}^{Z_c}$ (see Eq.(\ref{Hsquared})) and this approximation has less than  $1\%$ error in the physical region. A  similar observation was also made in Refs.\cite{Chen:2015jgl,Chen:2016mjn} based on the heavy-quark nonrelativisitic expansion.

The expression of the helicity amplitude $\mathcal{H}_{++}^{Z_c}$ in terms of the invariant amplitudes $F_i^{Z_c}(s,t)$ is given by
\begin{align}
    \mathcal{H}_{++}^{Z_c}(s,t) &=
    \frac{s-q^2-m_\psi^2}{2} \, F_1(s,t)- q^2  m_\psi^2 \, F_4(s,t)
    \nonumber \\
    &+(t-u)^2 \; \frac{ s (q^2+m_\psi^2) - (m_\psi^2 - q^2)^2}{2 \, \lambda(s,q^2,m_\psi^2)} \, F_2(s,t)
    \nonumber \\
    &+(t-u)^2 \; \frac{q^2 + m_\psi^2 }{2} \, F_3(s,t)\,,
\end{align}
where
\begin{align}\label{Eq:Fi(s,t)}
&    F_1^{Z_c} = - \frac{\mathcal{F}_{\gamma^* \pi Z }\, C_{Z \psi \pi}}{8} \left(
    \dfrac{4\,t + q^2 + m_{\psi}^2}{t - m_Z^2} + \dfrac{4\,u + q^2 + m_{\psi}^2}{u - m_Z^2}
    \right) ,
    \nonumber \\
&    F_2^{Z_c} = - \frac{\mathcal{F}_{\gamma^* \pi Z }\, C_{Z \psi \pi}}{8} \left(
    \dfrac{1}{t - m_Z^2} + \dfrac{1}{u - m_Z^2}
    \right),
 \\
&    F_3^{Z_c} =  \frac{\mathcal{F}_{\gamma^* \pi Z }\, C_{Z \psi \pi}}{4\,(t-u)} \left(
    \dfrac{1}{t - m_Z^2} - \dfrac{1}{u - m_Z^2}
    \right),
    \nonumber \\
&    F_4^{Z_c} = - \frac{\mathcal{F}_{\gamma^* \pi Z }\, C_{Z \psi \pi}}{4} \left(
    \dfrac{1}{t - m_Z^2} + \dfrac{1}{u - m_Z^2}
    \right),\nonumber\\    
&F_5^{Z_c} = 0\,.  \nonumber
\end{align}
Due to the polynomial ambiguity of the p.w. amplitudes, we will consider only the pole contribution. Based on the fixed-$s$ Mandelstam representation one can show that the pole contribution corresponds to fixing $t= m_Z^2$ and $u= m_Z^2$ in the numerators of Eq.(\ref{Eq:Fi(s,t)}). This procedure is in line with the definition of the on-shell transition form factor $\mathcal{F}_{\gamma^* \pi Z_{c}}(q^2)$ and does not change the amplitude in the physical region. 

\begin{figure}[t]
\begin{center}
\begin{overpic}[width=8cm]{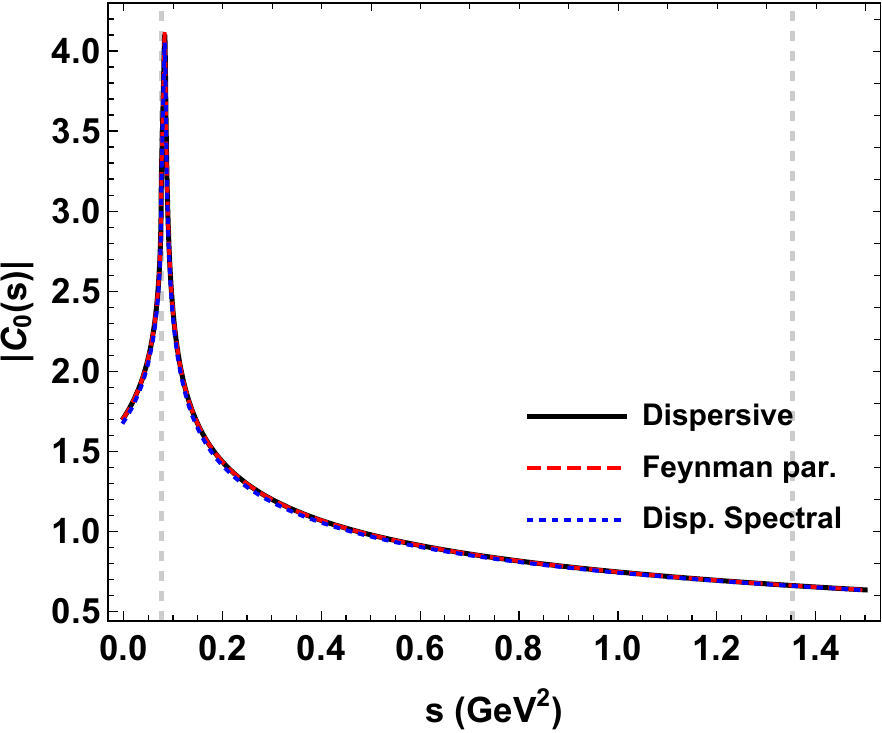}
 \put(30,40){\includegraphics[scale=.45]{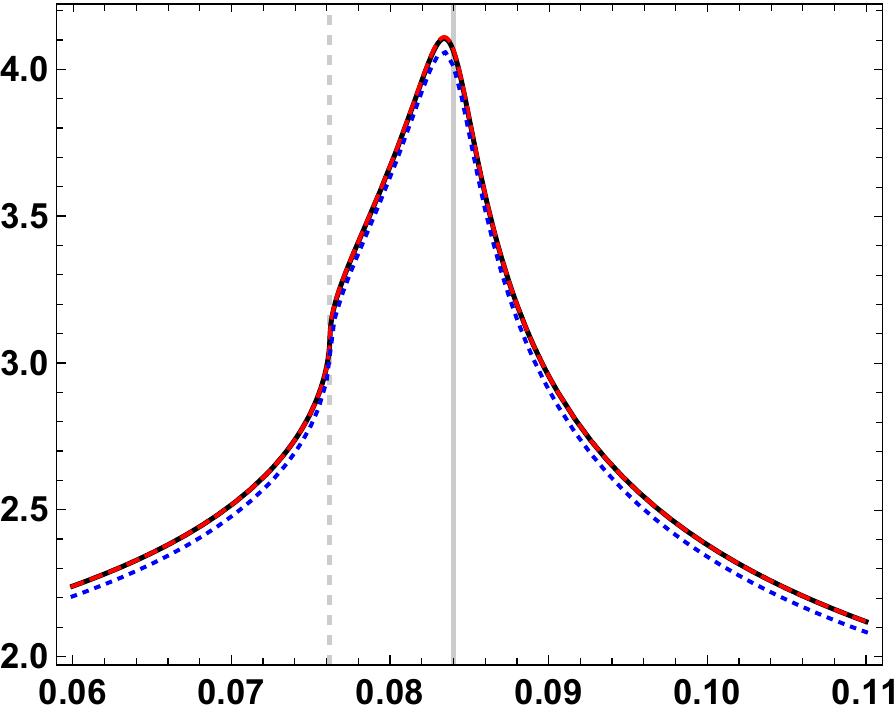}}
\end{overpic}
\end{center}
\caption{Comparison of the absolute values of the scalar triangle loop function $C_0(q^2,m_\psi^2,m_\pi^2,m_Z^2,m_\pi^2)$ calculated numerically using Feynman parameters (dashed red line) with a dispersive evaluation (solid black line) at $q=4.23$ GeV. The vertical dashed lines indicate the kinematically allowed decay region, while the solid gray line in the inset is exactly at the position of the triangle singularity.}
\label{triang}
\end{figure}

\subsection{Triangle Singularities} \label{triagle}
For the three-body decays, it is frequent that the left-hand cut overlaps with the right-hand cut and requires special treatment in the dispersive formalism. In our previous analysis of $\gamma^*(q^2) \to \psi(2S) \pi\pi$ \cite{Molnar:2019uos} such an overlap required a distortion of the integration path which was performed by including an additional anomalous piece \cite{Karplus:1958zz,Mandelstam:1960zz,Hoferichter:2013ama}. For the processes considered in the present paper, the overlap of the left and right-hand cuts does not introduce anomalous thresholds, but still require the proper analytical continuation for the energy variable $q^2 \to q^2 + i \epsilon$ \cite{Bronzan:1963mby,Moussallam:2013una} due to the presence of the so-called triangle singularity \cite{Guo:2019twa,Achasov:2015uua,Achasov:2018swa} associated with $Z_c \pi\pi$ loop. Indeed, for $q=4.23$ GeV and $q=4.26$ GeV the exchange of the $Z_c(3900)$ state in the triangle loop can be on-shell, satisfying the Coleman-Norton conditions $q^2 > (m_Z + m_\pi)^2$ and $m_Z^2 > (m_\psi + m_\pi)^2$ \cite{Coleman:1965xm}. This implies that the branch point $s_-$ associated with the left-hand cut
\begin{align}
s_{-} =& \frac{1}{2} \left[ 
q^2+m_{\psi}^2 + 2 m_{\pi}^2 - m_Z^2 - \dfrac{(q^2-m_\pi^2)(m_\psi^2-m_\pi^2)}{m_Z^2}
\right] \nonumber\\
&- \dfrac{\kappa(m_Z^2)}{2\, m_Z^2}\,,
\end{align}
is located just above the two pion threshold but infinitesimally below the real axis \cite{Szczepaniak:2015eza}. We note, that the $q^2 \to q^2 + i\, \epsilon$ continuation guarantees that the branching point never crosses the unitarity cut and the dispersive representations of Eqs.~(\ref{Eq:RescSwave}) and (\ref{Eq:RescDwave}) are correct. Due to the finite resonance width, however, the effect of the triangle singularity smears out, since the singular point is shifted further away from the physical region. 

There are several ways of accounting for the width of the $Z_c(3900)$ state. The proper implementation requires modeling the propagator using a spectral representation \cite{Moussallam:2013una}, i.e., it should have sound analyticity properties, such as pole on the unphysical Riemann sheet and the right-hand cuts starting at $\pi J/\psi$ and $D\bar{D}^*$ thresholds. This analysis is beyond the scope of our paper due to the lack of experimental information. Since the width of the $Z_c(3900)$ meson is relatively small ($\Gamma_Z=28.2$ MeV)  \cite{Tanabashi:2018oca}, we follow here a pragmatic approach by implementing the finite width in the denominators of Eq.(\ref{Eq:Fi(s,t)}). In this case, it is possible to cross-check our dispersive implementation on an example of a toy model of scalar fields with a constant (equal to unity) interaction between pions. As one can see in Fig.\ref{triang}, the result of the dispersive calculation and the calculation via Feynman parameters in perturbation theory give the same results. For illustrative purpose, we also show in Fig.\ref{triang} the result based on using the spectral representation of the $Z_c$ propagator \cite{Moussallam:2013una}, but accounting for just one channel $\pi J/\psi$ as it was done in Ref.\cite{Chen:2019mgp}. As expected the difference is negligible. Due to the narrowness of the $Z_c(3900)$ state one can also observe in Fig.\ref{triang} that the peak is still relatively sharp. However, the inclusion of the unitarization through the Muskhelishvili-Omn\`es representation  smears it out in the Dalitz plot.

\subsection{Omn\`es functions} \label{SubSec:hadronic_input}
For the S-wave isospin $I=0$ amplitude, we use the coupled-channel Omn\`es function from a dispersive summation scheme \cite{Gasparyan:2010xz, Danilkin:2010xd} which implements constraints from analyticity and unitarity. The method is based on the $N/D$ ansatz \cite{Chew:1960iv}, where the set of coupled-channel integral equations for the $N$-function are solved numerically with the input from the left-hand cuts which we present in a model-independent form as an expansion in a suitably constructed conformal mapping variable. These coefficients in principle can be matched to $\chi$PT at low energy \cite{Danilkin:2011fz,*Danilkin:2012ap}. Here we use a data-driven approach, and determine these coefficients directly from fitting to Roy analyses for $\pi\pi \to \pi\pi$ \cite{GarciaMartin:2011cn}, $\pi\pi \to K\bar{K}$  \cite{Buettiker:2003pp,*Pelaez:2018qny} and existing experimental data for these channels. After solving the linear integral equation for $N(s)$, the $D$-function (inverse of the Omn\`es function) is computed. The obtained coupled-channel Omn\`es function has already been successfully applied for the photon-fusion reactions $\gamma^{(*)}\gamma^{(*)}\to \pi\pi$ in \cite{Danilkin:2018qfn,Danilkin:2019mhd,Danilkin:2019opj,Deineka:2019bey}. The Omn\`es function for the D-wave ($I=0$) is constructed directly from the $\pi\pi$ phase shift \cite{GarciaMartin:2011cn} and given by
\begin{equation}\label{OmenesPhaseShift}
\Omega^{(2)}(s)=\exp\left(\frac{s}{\pi}\int_{4m_\pi^2}^{\infty} \frac{d s'}{s'}\frac{\delta_{I=0}^{(2)}(s')}{s'-s}\right)\,,
\end{equation}
since the inelasticity around $f_2(1270)$ peak is suppressed \cite{Tanabashi:2018oca}.

\begin{figure*}[!t]
\centering
\includegraphics[height=2.7cm]{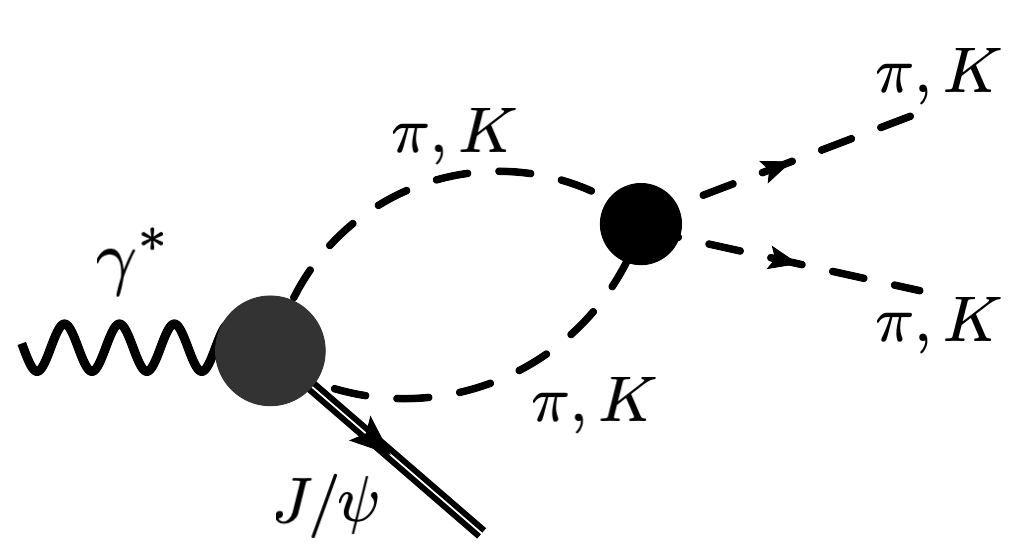} 
\caption{Diagrammatic representation of the final state interaction of the process $\gamma^* \to J/\psi \; \pi  \pi\,(K \bar{K})$.}
\label{fig-bubble}
\end{figure*}

\begin{figure*}[!t]
\centering
\includegraphics[height=2.7cm]{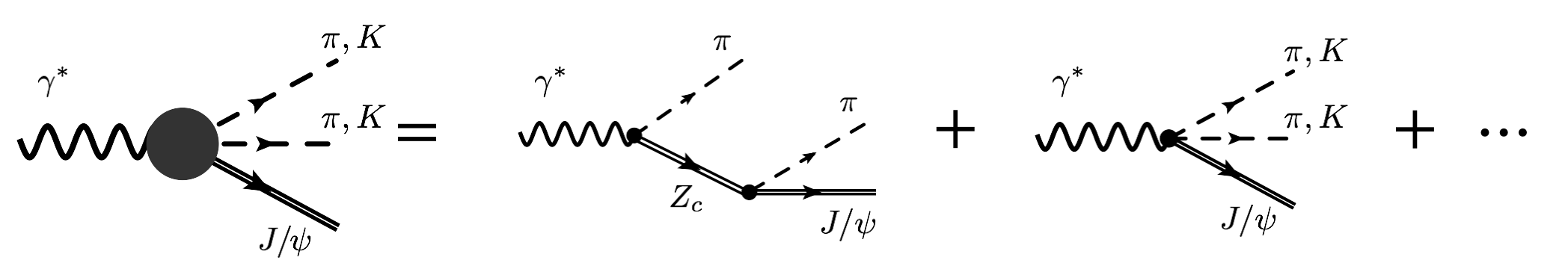} 
\caption{The physical meaning of the left vertex of Fig.\ref{fig-bubble}.
}
\label{diagrams}
\end{figure*}

\section{Results and Discussion} \label{results}

For the S-wave contribution we write a twice-subtracted dispersive representation. Due to the coupled-channel there are in total four subtraction constants. For the D-wave we allow for one subtraction. Even though the dispersive integrals are formally convergent with less subtractions\footnote{For the S-wave both $h^{(0),Z_c}_{++}(s')$ and $\text{Disc}\,  (\Omega^{(0)}(s'))^{-1}$ are asymptotically bounded (modulo slowly varying logarithm), while for the D-wave $h^{(2),Z_c}_{++}(s')/\gamma(s')_{s'\to \infty}\sim 1/s'^2$ and $\text{Disc}\,  (\Omega^{(2)}(s'))^{-1}_{s'\to \infty}\sim s'$.} they acquire significant corrections from the integration over large $s$. Therefore, we implement over-subtracted dispersion relations in order to reduce the sensitivity to the high energy region and the effects of additional unknown left-hand cuts, such as possible D-meson loops or contact interaction \cite{Chen:2019mgp, Chen:2016mjn}.
To check on the physical importance of the latter, we will also compare in the following for the S-wave contribution the fitted subtraction constants with the sum rule result which one would obtain from a once-subtracted dispersive formalism.

For the S- and D-waves we diagrammatically show the contributions in our formalism in Fig.\ref{fig-bubble} (with the input from Fig.\ref{diagrams}).
For all higher partial waves, we take the contribution of the pure $Z_c$ diagram only (first term on {\it rhs} of Fig.\ref{diagrams}).
Since the dispersion relations in Eqs. (\ref{Eq:RescSwave}) and (\ref{Eq:RescDwave}) are written for $I=0$ we need to encode the transformation coefficients between isospin and the physical amplitudes
\begin{align}
\mathcal{H}_{++}=\frac{1}{\sqrt{3}}\,\mathcal{H}_{0,++}\,,\quad \mathcal{K}_{++}=\frac{1}{\sqrt{2}}\,\mathcal{K}_{0,++}\,.
\end{align}
Therefore, for $e^+e^- \to J/ \psi \, \pi^+ \pi^-$ one obtains
\begin{align}\label{finalpipi}
&\mathcal{H}_{++} (s,t) = \frac{1}{\sqrt{3}}\, \Bigg[ \mathcal{H}^{Z_c}_{0,++}(s,t) 
  \\
& + \Omega_{11}^{(0)}\,\Bigg\{ 
a + b \, s 
-\dfrac{s^2}{\pi} 
 \int\limits_{4m_\pi^2}^{\infty} 
 \dfrac{d s'}{s'^2} \dfrac{\text{Disc}\, (\Omega^{(0)}(s'))^{-1}_{11}}{s' - s}h_{0,++}^{(0), Z_c}(s')
\Bigg\}
\nonumber  \\
& +   \Omega_{12}^{(0)}\,\Bigg\{ 
c + d \, s 
-\dfrac{s^2}{\pi} 
 \int\limits_{4m_\pi^2}^{\infty} 
 \dfrac{d s'}{s'^2} \dfrac{\text{Disc}\, (\Omega^{(0)}(s'))^{-1}_{21}}{s' - s} h_{0,++}^{(0), Z_c}(s')
\Bigg\}
\nonumber  \\
&+  5 \, P_{2}(z)\, \gamma(s) \, \Omega^{(2)}  \nonumber\\
&\qquad\quad \times \Bigg\{ e-\frac{s}{\pi}
 \int\limits_{4m_\pi^2}^{\infty} 
 \dfrac{d s'}{s'} \dfrac{\text{Disc}\,  (\Omega^{(2)}(s'))^{-1}}{s' - s} \dfrac{h_{0,++}^{(2), Z_c}(s')}{\gamma(s')}  \Bigg\} \Bigg], \nonumber
\end{align}
where $\mathcal{H}^{Z_c}_{0,++}(s,t)$ is a pure $Z_c$-exchange 
and we put $h^{(J),L}_{0,++}(s)=h^{(J),Z_c}_{0,++}(s)$ and $k^{(J),L}_{0,++}(s)=0$ according to the discussion given above. We note that the partial wave amplitudes $h_{0,++}^{(J),Z_c}(s)$ were properly modified due to the presence of logarithmic singularity (see section \ref{triagle}). For the $e^+e^- \to J/ \psi \, K^+K^-$ there is only a S-wave contribution corresponding to
\begin{align}\label{finalKK}
&\mathcal{K}_{++}(s,t)  =  
 \\
&\frac{\Omega_{21}^{(0)}}{2}\, \Bigg\{ 
a + b \, s 
-\dfrac{s^2}{\pi} 
 \int\limits_{4m_\pi^2}^{\infty} 
 \dfrac{d s'}{s'^2} \dfrac{\text{Disc}\, (\Omega^{(0)}(s'))^{-1}_{11}}{s' - s} \,h_{0,++}^{(0),Z_c}(s')
\Bigg\}
\nonumber  \\
&+ \frac{\Omega_{22}^{(0)}}{2}\, \Bigg\{
c + d \, s 
-\dfrac{s^2}{\pi} 
 \int\limits_{4m_\pi^2}^{\infty} 
 \dfrac{d s'}{s'^2} \dfrac{\text{Disc}\,  (\Omega^{(0)}(s'))^{-1}_{21}}{s' - s} \,h_{0,++}^{(0),Z_c}(s')
\Bigg\} \nonumber
\end{align}

\begin{figure*}[!t]
\centering
\includegraphics[width =0.475\textwidth]{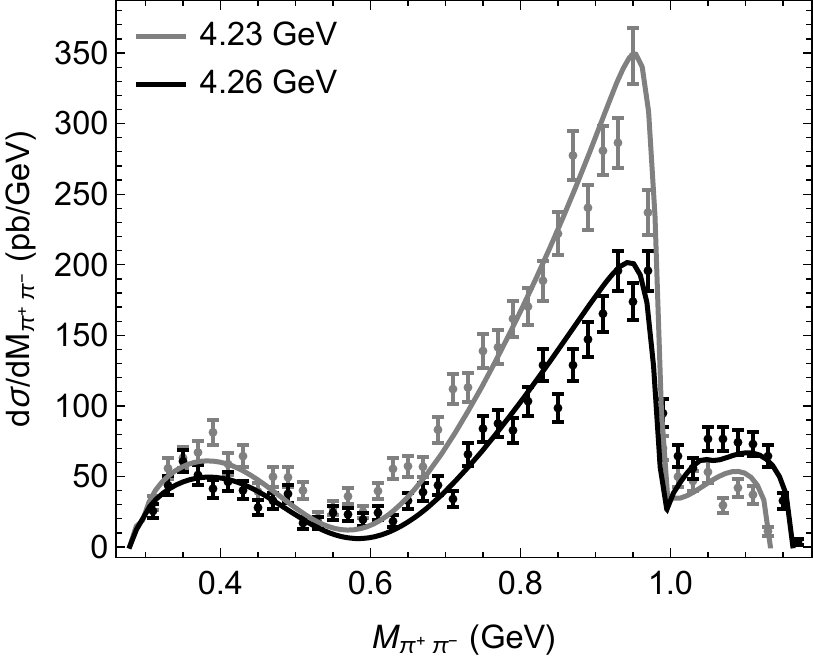}\quad 
\includegraphics[width =0.475\textwidth]{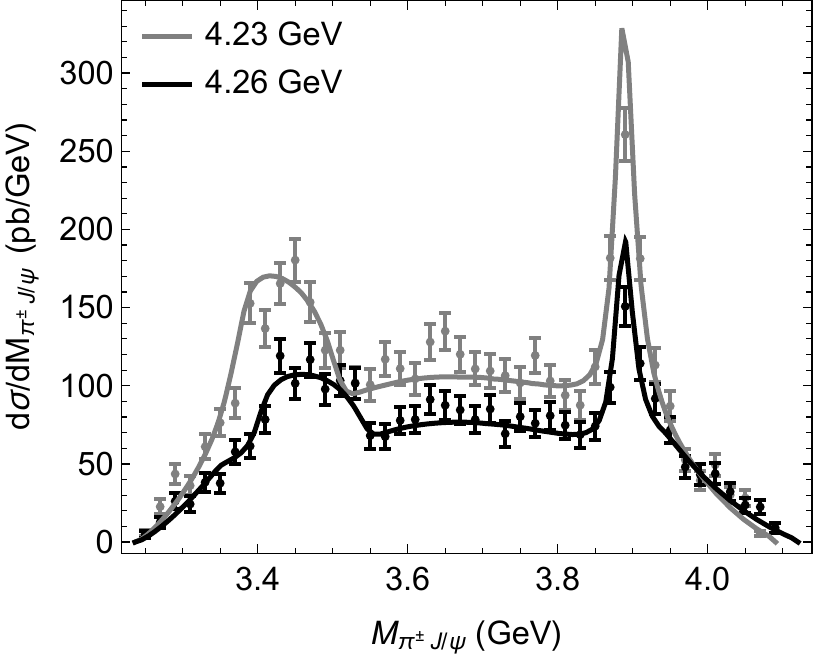}
\caption{The most economical fit with four real parameters at $q=4.23$ GeV and $q=4.26$ GeV (see Fit 1 in Table  \ref{Tab:q=4.23}). The BESIII data is taken from Ref.\cite{Collaboration:2017njt}, which was normalized to the total cross section given in Ref.\cite{Ablikim:2016qzw}. \label{Fig:4par}}
\end{figure*}

Below we perform a simultaneous fit to the $\pi^+\pi^-$ and $\pi^{\pm} J/\psi$ invariant mass distributions \cite{Collaboration:2017njt} together with the total cross-section data for  $\sigma(J/\psi K^+K^- )$ \cite{Ablikim:2018epj}. To ensure that the $e^+e^-\to J/\psi K^+K^- $ total cross-section constraint is well accounted for and contributes realistically to the total $\chi^2$, we re-scale its error by the amount of experimental data points above the $K\bar{K}$ threshold in the $\pi\pi$ distributions. In our fits we therefore minimize 
\begin{equation}
\chi^2_{tot}=\frac{1}{N_{dof}}\left(\chi^2_{\pi\pi}+\chi^2_{\pi\psi}+\chi^2_{KK}\right)\,,
\end{equation}
where 
\begin{eqnarray}\label{Eq:different_chi^2}
\chi^2_{\pi\pi}&=&\sum_{i=1}^{N_{\pi\pi}} \left(\frac{d\sigma^{\text{Th}}_i/d M_{\pi\pi}- d\sigma^{\text{Exp}}_i/d M_{\pi\pi}}{\Delta(d\sigma^{\text{Exp}}_i/d M_{\pi\pi})}\right)^2, \nonumber\\
\chi^2_{\pi\psi}&=&\sum_{i=1}^{N_{\pi\psi}} \left(\frac{d\sigma^{\text{Th}}_i/d M_{\pi\psi}- d\sigma^{\text{Exp}}_i/d M_{\pi\psi}}{\Delta(d\sigma^{\text{Exp}}_i/d M_{\pi\psi})}\right)^2,\\
\chi^2_{KK}&=&\left(\frac{\sigma(J/\psi K^+K^-)^{\text{Th}}-\sigma(J/\psi K^+K^-)^{\text{Exp}}}{\Delta \sigma(J/\psi K^+K^-)^{\text{Exp}}/\sqrt{2\,N_{KK}}}\right)^2\,, \nonumber
\end{eqnarray}
with 
\begin{align}\label{Eq:Ndof}
N_{dof}=N_{\pi\pi}+N_{\pi\psi}+2\,N_{KK}-N_{\text{par}}\,.
\end{align}
The number of data-points are: $\{N_{\pi\pi},N_{\pi\psi},N_{KK}\}$ $=\{42,\,42,\,7\}$ for $q=4.23$ GeV and $\{N_{\pi\pi},N_{\pi\psi},N_{KK}\}=\{44,\,43,\,8\}$ for  $q=4.26$ GeV. 
Note that in the $\pi\pi$ and $\pi\psi$ data sets we omitted the bins that cross the boundary of the Dalitz plot. 

\begin{table*}[t]
\renewcommand*{\arraystretch}{1.4}
\begin{tabular*}{\textwidth}{@{\extracolsep{\fill}}|l|c|c|c|c|@{}}
\hline 
& \multicolumn{2}{c|}{$q=4.23$ GeV}   & \multicolumn{2}{c|}{$q=4.26$ GeV} \\
\hline  &  Fit 1 & Fit 2  & Fit 1 & Fit 2  \\
\hline
$|e\, \mathcal{F}_{\gamma^* \pi Z }\, C_{Z \psi \pi} |^2\times 10^{7}$   
                          &  5.8(4)  & 3.4(3)   & 2.9(2)      & 1.3(2) \\
$\tilde{a}\times 10^{-3}$ &  3.3(1)  & 3.9(2)   & 4.1(2)      & 5.3(4) \\
$\phi_{a} (\mathrm{rad})$ &  $-$     & -0.50(2) & $-$        & -0.33(2) \\
$\tilde{b}\times 10^{-3}$ &  -9.2(4) & -11.2(6) & -11.2(5)  & -15.8(1.2) \\
$\phi_{b} (\mathrm{rad})$ &  $-$     & -0.20(2) & $-$          & $-$ \\
$\tilde{c}\times 10^{-3}$ &  $-$     & $-$      & $-$         & 4.6(6) \\
$\tilde{d}\times 10^{-3}$ &  -4.0(1) & -5.0(3)  & -4.3(2)    & -11.6(1.0) \\
\hline
$\tilde{e}\times 10^{-2}$ &  fixed to sum rule   & 8.1(1.1)   & fixed to sum rule  & 3.1(2.5) \\
\hline
$\sigma(J/\psi K^+K^-)^{\text{Exp}}$  [pb] & \multicolumn{2}{c|}{$5.3(1.0)$}   & \multicolumn{2}{c|}{$3.1(6)$} \\
\hline
$\sigma(J/\psi K^+K^-)^{\text{Th}}$  [pb]  
                          & 4.4(5)  & 5.2(2) & 2.9(4)  & 3.0(3) \\
\hline
$\chi_{\text{tot}}^2$ &  3.4     & 1.7  &  2.5  &   1.3\\
\hline
\end{tabular*}
\caption{Fit parameters entering Eqs.(\ref{finalpipi}) and (\ref{finalKK}) which were adjusted to reproduce the empirical $\pi\pi$ and $\pi J/\psi$ invariant mass distributions together with the cross-section $\sigma(J/\psi K^+K^-)$ at $e^+e^-$ center-of-mass energies $q=4.23$ GeV and $q=4.26$ GeV. Tildes on top of a subtraction constants indicate that they are given relative to the couplings constants entering $h_{0,++}^{(J),Z_c}$, for instance $\tilde{a} \equiv a/(\mathcal{F}_{\gamma^* \pi Z }\,C_{Z \psi \pi})$. For easier comparison of the fits with real subtraction constants ($\phi_i=0$) and the fits with complex subtraction constants  ($\phi_i \neq0$), we restricted $\phi_{i}$ in the region $(-\pi/2,\pi/2)$, i.e. allowing to have $\pm$ signs in front of the absolute value. Errors on fit parameters are shown in brackets. 
\label{Tab:q=4.23}}
\end{table*}

Due to an overlap of left- and right-hand cuts, the subtraction constants ($a,b,c,d,e$) can in principle be complex, which together with the product $\mathcal{F}_{\gamma^* \pi Z }\,C_{Z \psi \pi}$ leaves us with eleven parameters for each $e^+e^-$ center-of-mass energy to describe the data. We definitely do not want to over-fit the data and describe some variations in the data that could just be statistical noise. Therefore, we decided to start with the most economical fit in which we fit four parameters as described in the following, and will then compare it with our best fit which has seven parameters. The summary of the fit results is given in Table \ref{Tab:q=4.23}.

\begin{figure*}[!t]
\centering
\includegraphics[width =0.475\textwidth]{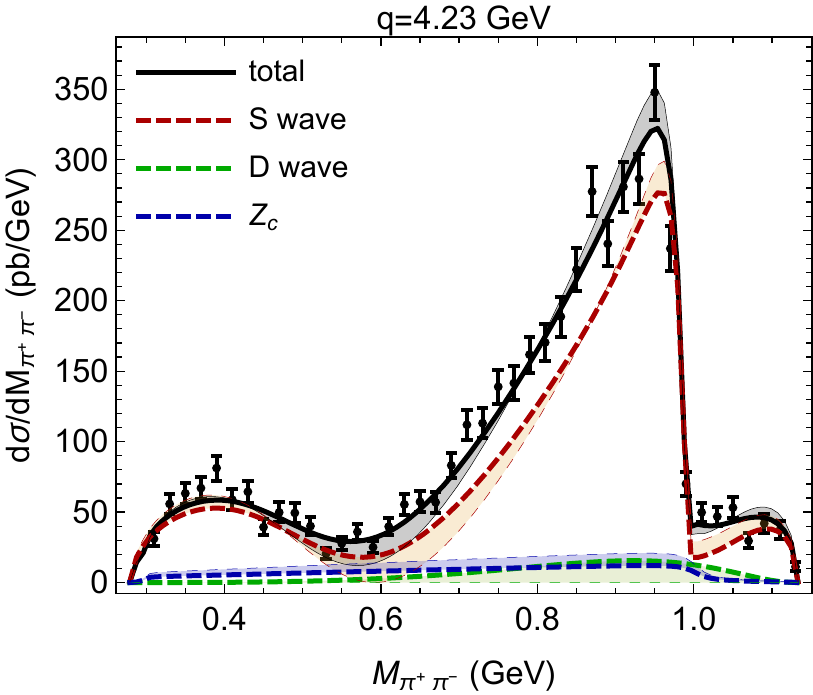}\quad 
\includegraphics[width =0.475\textwidth]{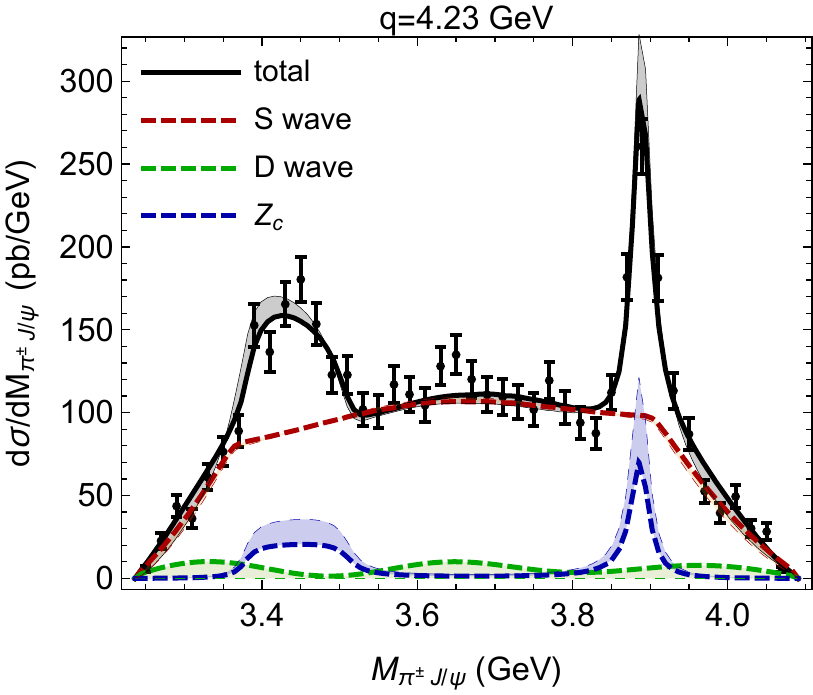}\\
\vspace{0.5cm}
\includegraphics[width =0.475\textwidth]{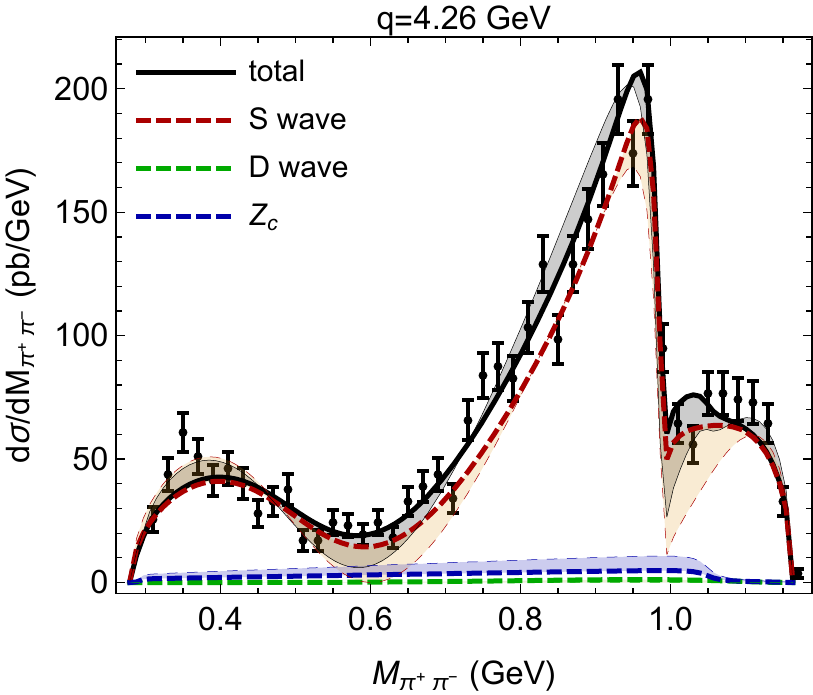}\quad 
\includegraphics[width =0.475\textwidth]{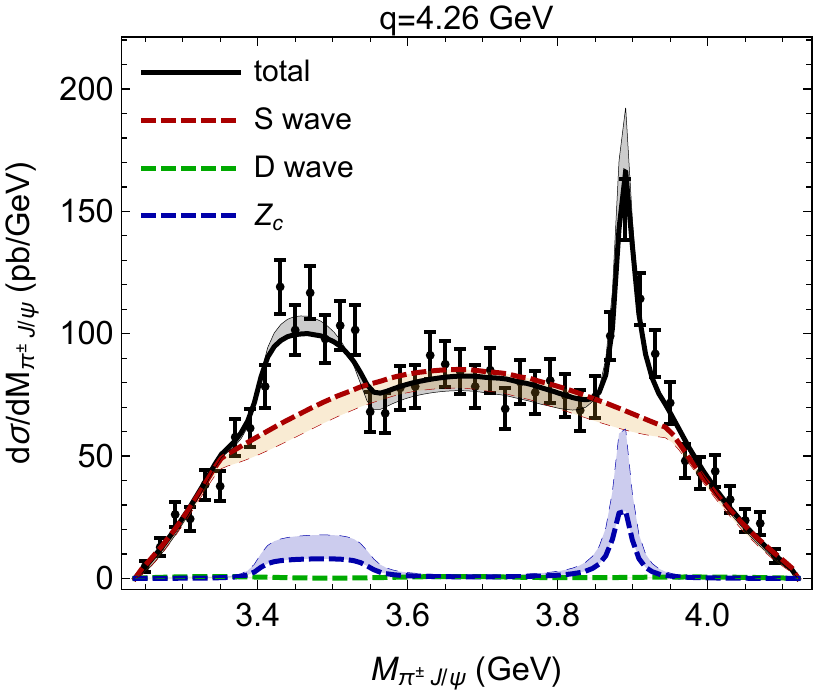}\\
\caption{The black curves are the total fit results. The individual contributions from the pure $Z_c(3900)$-exchange, the $\pi\pi$ re-scattering in the S and D waves are indicated by blue, red and green curves, respectively. On a practical level, these curves correspond to the first, second plus third and forth term of Eq.(\ref{finalpipi}), as explained in the text.
The shaded bands indicate the spread between Fit 1 (thin curves) and Fit 2 (thick curves) results. The BESIII data is taken from Ref.\cite{Collaboration:2017njt}, which was normalized to the total cross section given in Ref.\cite{Ablikim:2016qzw}. \label{Fig:7par}}
\end{figure*}

We start with the case when all the subtraction constants are real in the S-wave while for the D-wave we use an unsubtracted dispersive representation. It turns out that the fitted value of the $c$ parameter is consistent with zero and therefore it is justified to ignore it for this initial fit. This leaves us with four real parameters. Even though this parameterization is not perfect, it provides a good description of the data as shown in Fig.\ref{Fig:4par}. In particular, in the $\pi\pi$ mass distribution the dip structure around the $K\bar{K}$ threshold comes out naturally in our formalism due to the $f_0(980)$ resonance. In addition, the PDG \cite{Tanabashi:2018oca} averaged mass and the width of $Z_c(3900)$,  $m_Z=3.8872(23)$ GeV and $\Gamma_Z=28.2(2.6)$ MeV, seem to be well in agreement with the data for the $\pi\psi$ mass distribution. Furthermore, it is worth mentioning an interesting observation: if we fit only the $\pi\psi$ invariant mass distribution for $q=4.23$ GeV or $q=4.26$ GeV, the post-diction for the $\pi\pi$ distribution reproduces very well the major features of the data\footnote{The opposite is not true, because by fitting only the $\pi\pi$ distribution it is hard to constrain well the parameters of the $Z_c$ state  and the post-diction of the $\pi\psi$ distribution is then only qualitative.}. This implies that our framework has the correct ingredients in the simultaneous description of the data. As seen from the parameter values of Fit 1 in Table \ref{Tab:q=4.23}, we  also find that they  not vary much between $q=4.23$ GeV and $q=4.26$ GeV. This is in accordance with our expectation since the considered $e^+e^-$ center-of-mass energies are different only by $30$ MeV. Therefore, the parameters of Fit 1 determine the starting values of our improved fit. 

A significant improvement over Fit 1 can be obtained by adding a phase to the parameter $a$ and to a lesser extent also to the parameter $b$, since the subtraction constants $a$ and $b$ are mainly responsible for the description of the data below the $K\bar{K}$ threshold. The region above $K\bar{K}$ threshold is a bit more complicated since the $c$ and $d$ parameters play a significant role there. Due to the absence of the $K\bar{K}$ mass distribution data, we keep the subtraction constants $c$ and $d$ real. From the analysis of different fits we found that for $q=4.23$ GeV a small non-zero value of the phase $\phi_b$ allows to improve the fit more, while for $q=4.26$ GeV the parameter $c$ plays a more prominent role. In addition, we allow for one subtraction in the D-wave, which may differ from the unsubtracted sum rule value. As a result, we decided to limit ourselves to the following 
``best" fit scenario with seven parameters: for $q=4.23$ GeV we consider the product $\mathcal{F}_{\gamma^* \pi Z }\,C_{Z \psi \pi}, a, \phi_a, b, \phi_b, d$, and $e$ as fit parameters, while for $q=4.26$ GeV we consider the product $\mathcal{F}_{\gamma^* \pi Z }\,C_{Z \psi \pi}, a, \phi_a, b, c, d$, and $e$ as fit parameters. 

\begin{figure*}[t]
\centering
\includegraphics[width =0.475\textwidth]{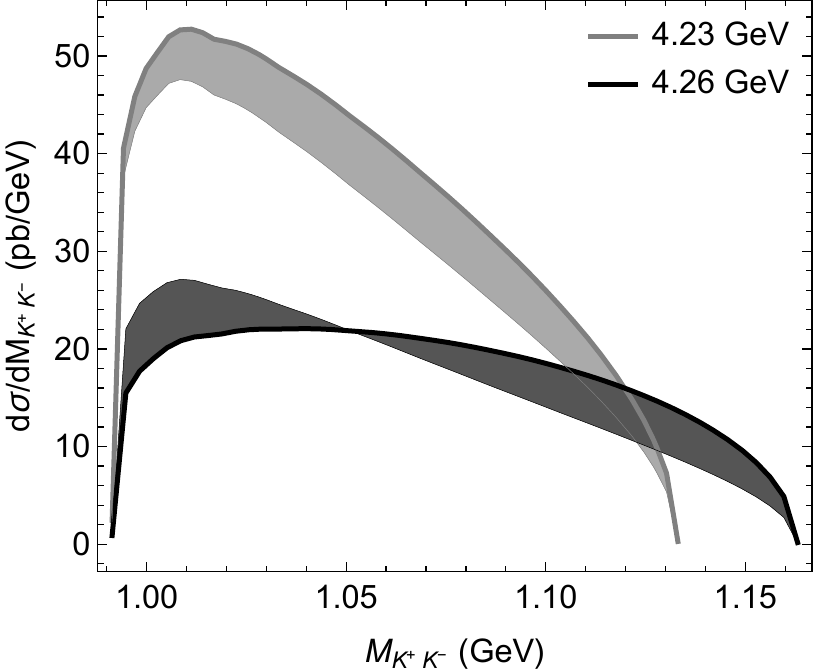}\quad 
\includegraphics[width =0.475\textwidth]{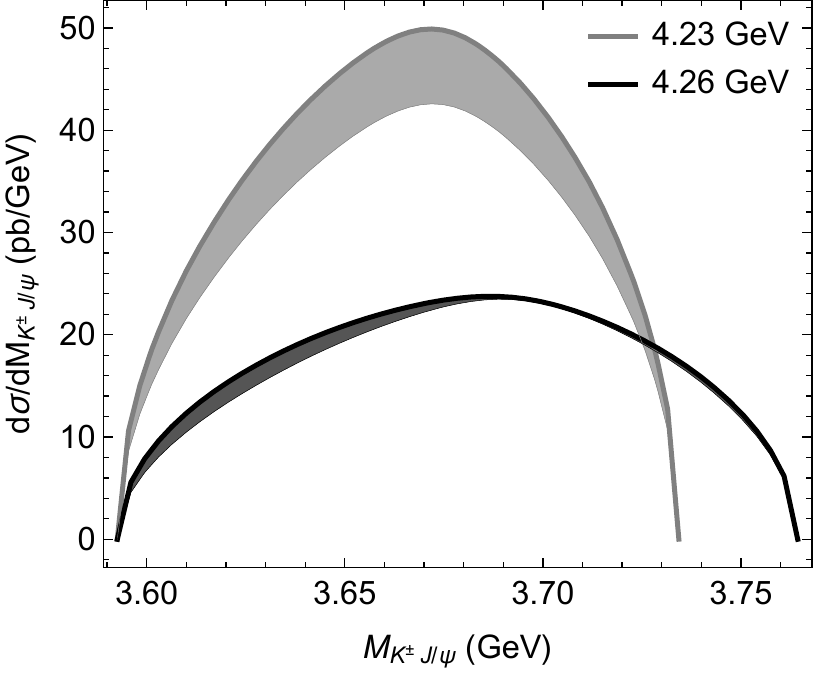}
\caption{Theoretical predictions of the $K^+K^-$ and $K^{\pm}J/\psi$ invariant mass distributions of the $e^+e^- \to J/\psi K^+K^-$ reaction for $e^+e^-$ center-of-mass energies $q=4.23$ and $q=4.26$ GeV. The shaded bands indicate the spread between Fit 1 (thin curves) and Fit 2 (thick curves) results. \label{Fig:prediction}}
\end{figure*}

The resulting parameters and $\chi^2$ are collected under the Fit 2 in Table \ref{Tab:q=4.23} and shown in Fig.\ref{Fig:7par}, where we also show contributions from the individual terms in Eq.(\ref{finalpipi})
.
We see that our results are in very good agreement with the data. As a conservative error estimate we show in Fig.\ref{Fig:7par} the spread between the Fit 1 (our most economical fit) and Fit 2 (our best fit) results. We found that the remaining parameters, beyond the seven parameters considered, have a rather small effect on the $\pi\pi$ and $\pi\psi$ distributions and can be determined only when very precise data will be available.

It is instructive to compare the fitted values of the $\tilde{b},\,\tilde{d}$ and $\tilde{e}$ parameters from Table \ref{Tab:q=4.23} with the sum rule (SR) estimates resulting from a once-less subtracted  dispersion relation. Such framework implies e.g. for the parameter $b$ the relation 
\begin{align}
b^{SR}=-
 \int\limits_{4m_\pi^2}^{\infty} 
\dfrac{d s'}{\pi} \dfrac{\text{Disc}\,  (\Omega^{(0)}(s'))^{-1}_{21}}{s'^2} \,h_{0,++}^{(0),Z_c}(s'),
\end{align}
and analogous relations for $d$ and $e$. 
Using our $Z_c$ pole model for the left-hand cut we obtain
$$
\renewcommand*{\arraystretch}{1.4}
\begin{tabular*}{\columnwidth}{@{\extracolsep{\fill}}lll}
\hline   & $q=4.23$ GeV & $q=4.26$ GeV \\
\hline
$(\tilde{b}^{\text{SR}}\times10^{-3},\phi_b)$ & (-0.6,0.9) & (-0.6,1.0)  \\
$(\tilde{d}^{\text{SR}}\times10^{-3},\phi_d)$ & (-0.3,1.0) & (-0.3,1.0) \\
$(\tilde{e}^{\text{SR}}\times10^{-2},\phi_e)$ & (-0.07,-0.9) & (-0.06,-0.7)\\
\hline
\end{tabular*}
$$
which for $\tilde{b}$ and $\tilde{d}$ are approximately 20 times smaller in magnitude (and even more for $\tilde{e}$) than the fitted values. This implies that besides the direct production of the $Z_c$ (first term in Eq.\eqref{finalpipi}), which is responsible for the peak regions in the $\pi \psi$ distribution, the two pions are predominantly produced directly in the transition from the $Y$ state to the $J/\psi$ state through a contact term and subsequently rescatter. Our analysis thus shows that the rescattering of the two pions happens  predominantly  without going through the  $Z_c(3900)$ state.  
In comparison, the dispersive analysis in Ref.\cite{Chen:2019mgp} indicates that the left-hand cut contributions from $Z_c$ are as significant as the chiral contact interaction and the D-wave contribution is comparable to the S-wave contribution in almost the whole phase space. Apart from a different treatment of the D-wave rescattering in a dispersive formalism (as discussed following Eq.(\ref{Eq:RescDwave})), it is hard to compare both approaches since we do not imply any particular dynamics on the contact interaction. The main aim of the present work is to perform a data-driven analysis of both Dalitz projections, in contrast to Ref.\cite{Chen:2019mgp}.

Since we obtained a simultaneous and accurate description of the BESIII data for the $\pi^+\pi^-$ and $\pi^{\pm}J/\psi$ invariant mass distributions, we find it justified to predict the $K^+K^-$ mass distribution. As one can see in Fig.\ref{Fig:prediction}, the obtained shape has a rapid rise just above the threshold, which is quite different from the pure phase space,  i.e. when $\mathcal{K}_{++}(s,t)$ is replaced by a constant. This behavior is due to $f_0(980)$ resonance and we expect to see it in future experimental measurements. For completeness, we also provide the prediction to $K^{\pm}J/\psi$ mass distribution, which is just a pure phase space in our approximation.

\section{Summary} \label{Sec:Summary}

In this work, we provided a quantitative and simultaneous description of the $\pi^+\pi^-$ and $\pi^{\pm} J/\psi$ invariant mass distributions of the recent BESIII data on  $e^+ e^- \to J/\psi \; \pi^+ \pi^-$ together with the total cross-sections $\sigma(J/\psi K^+K^-)$ at $e^+e^-$ center-of-mass energies $q=4.23$ GeV and $q=4.26$ GeV. A crucial element of our analysis is the well established charged exotic state $Z_c(3900)$, which we account for explicitly in $t$- and $u$-channels. The final state interaction of the two pions in the S- and D-waves is treated using the dispersion theory. For the S-wave, we consider coupled-channel unitarity since the kinematical region goes beyond the inelastic channel $K\bar{K}$ and the effect from the $f_0(980)$ resonance impacts  significantly the observables. On the other hand, for the D-wave a single-channel Omn\`es approach is adopted since the lowest resonance in that channel, the $f_2(1270)$ tensor resonance, decays predominantly into two pions. The final amplitudes depend on a set of  subtraction constants, which have been fitted to the BESIII data. A simultaneous description of $\pi^+\pi^-$ and $\pi^{\pm} J/\psi$ mass distributions together with the cross-sections $\sigma( J/\psi K^+K^-)$ is achieved through a four-parameter fit. We showed that the latter can be further improved by adding phases to the subtraction constant and allowing for one subtraction in the D-wave contribution. We found that the resulting seven parameter fit yields a very good description of the $\pi^+\pi^-$ and $\pi^{\pm} J/\psi$ mass distributions together with the cross-sections $\sigma( J/\psi K^+K^-)$. 
Our dispersive formalism shows that besides the direct production of the $Z_c$, responsible for the peak regions in the $\pi \psi$ distributions, the two pions are predominantly produced through a contact term in the transition from the $Y$ state to the $J/\psi$ state and subsequently rescatter.
For the $e^+ e^- \to J/\psi\, K \bar{K}$ we provided the first theoretical prediction for the two-kaon invariant mass distribution, which is significantly different from the pure phase space.

The constructed amplitudes provide powerful tools to analyze future data by the BESIII and Belle II  Collaborations. It can also be readily applied to study $e^+e^-$ annihilation into $\Upsilon(nS)\pi^+\pi^-$, where charged bottomonia like $Z_b^{\pm}$ states have been observed.

\section*{Acknowledgements}

The authors acknowledge Zhiqing Liu and Achim Denig for kindly providing the Dalitz projections. This work was supported by the Deutsche Forschungsgemeinschaft (DFG, German Research Foundation), in part through the Collaborative Research Center [The Low-Energy Frontier of the Standard Model, Projektnummer 204404729 - SFB 1044], and in part through the Cluster of Excellence [Precision Physics, Fundamental Interactions, and Structure of Matter] (PRISMA$^+$ EXC 2118/1) within the German Excellence Strategy (Project ID 39083149).

\bibliographystyle{apsrevM}
\bibliography{Bibliography}

\end{document}